\documentclass[copyright]{eptcs}
\providecommand{\event}{QAPL 2010} 
\providecommand{\volume}{??}
\providecommand{\anno}{2010}
\providecommand{\firstpage}{1}
\providecommand{\eid}{??}
\usepackage{breakurl}        

\usepackage[T1]{fontenc}
\usepackage[english]{babel}
\usepackage[latin1]{inputenc}

\usepackage{latexsym}
\usepackage{amsmath}
\usepackage{amssymb}
\usepackage{amsfonts}

\usepackage{pslatex}
\usepackage{theorems}

\usepackage{subfigure}

\usepackage[all]{xy}
\xymatrixrowsep{.7pc}

\usepackage{tikz}
\usetikzlibrary{arrows,shapes}

\tikzstyle{p1}=[circle, draw=black, semithick]
\tikzstyle{p2}=[rectangle, draw=black, semithick]

\usepackage{listings}
\def\myeps{\varepsilon}

\def\vini{v_\mathrm{ini}}
\def\limit{L}
\def\efrom#1{_{#1}E}
\def\eto#1{E_{#1}}
\def\S-1k{{[k-1]}}
\def\Sk{{[k]}}

\def\acolor{a}
\def\bcolor{b}

\def\diff{{\mathit{diff}}}
\def\pat{\rho}

\newcommand{\zeroMatrix}{\mathbf{0}}

\def\bigo{{\cal O}}
\newcommand{\nats}{\mathbb{N}}

\newcommand{\reals}{\mathbb{R}}
\newcommand{\rats}{\mathbb{Q}}
\newcommand{\ints}{\mathbb{Z}}

\def\int#1{\semb{#1}}

\def\intv#1{[#1]}
\def\neg#1{\overline{#1}}
\def\unknown#1{\widetilde{#1}}

\newcommand{\qdot}{\,.\,}

\newcommand{\conp}{{Co-NP}}
\newcommand{\shuffle}[2]{#1 \otimes #2}

\title{Quantitative Fairness Games\footnote{Work supported by MIUR PRIN Project 2007-9E5KM8.}
}

\author{Alessandro Bianco \and Marco Faella \and
Fabio Mogavero \and Aniello Murano
\institute{Universit\`a degli Studi di Napoli "Federico II", Italy}
}

\def\titlerunning{Quantitative Fairness Games}
\def\authorrunning{Bianco, Faella, Mogavero, and Murano}
\begin{document}


\maketitle

\begin{abstract}
We consider two-player games played on finite colored graphs where the goal is
the construction of an infinite path with one of the following frequency-related
properties: \emph{(i)} all colors occur with the same asymptotic frequency,
\emph{(ii)} there is a constant that bounds the difference between the
occurrences of any two colors for all prefixes of the path, or \emph{(iii)} all
colors occur with a fixed asymptotic frequency.
These properties can be viewed as quantitative refinements of the classical
notion of fair path in a concurrent system, whose simplest form checks whether
all colors occur infinitely often. In particular, the first two properties
enforce equal treatment of all the jobs involved in the system, while the third
one represents a way to assign a given priority to each job.
For all the above goals, we show that the problem of checking whether there
exists a winning strategy is CoNP-complete.
\end{abstract}


%



\section{Introduction}

Colored graphs, which are graphs with color-labeled edges,
are a model widely used in the field
of computer science that deals with the analysis of concurrent systems~\cite{MPvol1}.
%
For example, they can represent the transition relation of a concurrent program.
In this case, the color of an edge indicates which process
is making progress along that edge.
One basic property of interest for these applications is \emph{fairness}.
This property essentially states that, during an infinite computation,
each process is allowed to make progress infinitely often~\cite{Francez86}.
Starting from this core idea, a rich theory of fairness has been developed,
as witnessed by the amount of literature
devoted to the subject (see, for instance, \cite{luca-fair99,Kwiat89,LPS81}).

In the abstract framework of colored graphs,
the above basic version of fairness asks that, along an infinite path in the graph,
each color occurs infinitely often.
Such a requirement does not put any bound on the amount of steps that a process
needs to wait before it is allowed to make progress.
As a consequence, the asymptotic frequency of some color could be zero,
even if the path is fair.
Accordingly, several authors have proposed stronger
versions of fairness.
For instance, Alur and Henzinger define \emph{finitary fairness} roughly
as the property requiring that there is a fixed bound on the number of steps between two
occurrences of any given color~\cite{AH98,CH06}.
A similar proposal, supported by a corresponding temporal logic, was made by
Dershowitz, Jayasimha, and Park in~\cite{DJP03}.
On a finitarily fair path, all colors have positive asymptotic frequency.
%
These definitions of fairness treat the frequencies of the relevant events
in isolation and in a strictly qualitative manner.
Such definitions only distinguish between
zero frequency (not fair), limit-zero frequency (fair, but not finitarily so),
and positive frequency (finitarily fair).
Recently, we presented two new notions of fairness that introduce
a quantitative comparison between competing events~\cite{BFMM09}.
The \emph{balanced} path-property requires that on the path all colors occur
with the same asymptotic frequency, i.e.,
the long-run average number of occurrences for each of them is the same.
The \emph{bounded difference} path-property is a stronger property,
namely it requires that there is a numerical constant bounding the difference between the number
of occurrences of any two colors, for all prefixes of the path.
These notions provide stronger criteria suitable
for scheduling applications based on a coarse-grained model of the jobs involved.
In \cite{BFMM09}, by using a reduction to the feasibility of a linear system,
we proved that the problem asking whether there exists in a colored graph a balanced or a bounded
path is solvable in polynomial time.
%


A natural extension along this line of research is the introduction of a second decision
agent in the system, thus switching from graphs to games.
Games are widely used in computer science as models to describe
the interaction between a system and its environment \cite{KVW01,McN93,Tho95,Zie98}.
Usually the system is a component that is under the control
of its designer and the environment represents all the components the designer
has no direct control of. In this context, a game allows the designer to easily
check whether the system has the possibility to force some desired behavior
(or to avoid an undesired one) independently of the choices of the other external components.
A game comprises a graph that models the interaction between the entities involved,
commonly called \emph{players}. In this graph, a node represents a state of the
interaction, and an edge represents a progress in the interaction. We consider games where each state
is associated to only one player, and this player is the only one to have the possibility to choose
the progress toward a next state. A sequence of edges of the graph represents
a run of the system. Each player wants to force some runs with a desired property,
and it is said that he can \emph{win} the game if he can force a run with that property
independently of the choices of the other player. In this context, a \emph{strategy}
for a player is a predetermined decision that the player makes on all possible finite
paths ending with a node associated to that player.
%


%
In this paper, we address and study \emph{two-player colored games}, i.e., games where the underlying graph is a colored graph and the game is played between two players, which we refer to as \emph{player $0$} and \emph{player $1$}. In particular, we focus on
the goal for player $0$ to construct a balanced or a bounded path.
%
%
We believe that this game model can be useful in several formal verification contexts.
Coming back to the scheduling application, it can be useful in the case the scheduler may want to allow
a certain degree of freedom on the choices of lengthy
jobs that have to be executed by some components. More specifically, assume that due to
a design issue, the main scheduler can decide which macro-operation has to be executed and
then, some other schedulers can take decisions regarding some sub-operations of
the selected macro-operation.
In this context, our game model allows to check if the main scheduler has the
ability to force a balanced or a bounded progress of the activities, independently
of the sub-choices of the other schedulers.
As a specific example, consider the problem of synthesizing a fair scheduler
for a given set of concurrent jobs with shared resources \cite{dAFMR05}.
Assume that the jobs are known as data-abstract control-flow graphs.
Then, the resulting problem can be modeled as a two-player game between
the scheduler and the internal non-determinism of each job.
The scheduler (player $0$) tries to choose a correct sequence of jobs satisfying one of the two criteria discussed above, regardless of the non-deterministic choices made by the jobs (i.e., the moves of
player $1$).
Our main result shows that, in a game where the goal of player $0$
is the construction of a balanced or bounded path,
the problem of asking whether this player can always force such a path is \conp-complete.
For the lower bound, we use a reduction from the validity problem for boolean formulas.
For the upper bound, we first show that, in our game setting,
if player $1$ has a winning strategy, then he has a memoryless winning strategy.
Using this property, we decide whether there exists a winning strategy for player $0$
by simply checking whether all memoryless strategies for player $1$ are non-winning.
For a memoryless strategy of player $1$, we prune the game graph in accordance with the strategy and check whether, on the resulting subgraph, there exists a path satisfying the desired goal.
Such a path does not exist iff that strategy is winning for player $1$.
In the end, by guessing which memoryless strategy for player $1$ is winning,
we obtain a \conp\ algorithm that determines whether or not there exists a winning strategy for player $0$.

Sometimes, systems require that some jobs are executed more often than others.
In such a situation, it is useful to associate to each job a ``priority'' representing
how often that job should be executed compared to the others. \emph{Priority scheduling}
is a problem widely studied in computer science \cite{MM93}, usually with the
objective of minimizing the execution time of a given computation.
In general, a priority scheduling problem is NP-hard \cite{MM93}
and becomes solvable in polynomial-time if there are some restrictions on the nature of the system \cite{CD73}.
In this paper, we address and solve a new scheduling problem for a system
characterized by a finite number of states and infinite computation. As before, the system is modeled by a colored graph, where each color is associated with a given job.
We are interested in an execution of the system that spends a determined amount of time on each job. In our framework, the problem translates in looking for a path where each color occurs with some \emph{fixed asymptotic frequency}.
We call such a path a \emph{frequency path}.
%
We investigate this problem both in the (two-player) game and non-game setting.
In the game setting, the problem precisely consists of checking whether player $0$
can always force the construction of a frequency path (\emph{frequency goal}).
By using an argument similar to that used for games with balanced and bounded goals, we show that also games with frequency goals are \conp-complete.
In the non-game setting, by using a reduction to the feasibility of a linear system,
we show that the problem is much easier and solvable in polynomial-time.

\paragraph{Overview.}
The rest of the paper is organized as follows.
In Section~\ref{sec:defs}, we introduce some preliminary notation.
In Section~\ref{sec:balgames}, we introduce colored games with balanced,
bounded, and frequency goals and show that in all goal cases, the problem of deciding whether player $0$ has a winning strategy starting from a given node of the graph is \conp-complete.
In Section~\ref{sec:graphs}, we consider the (non-game) problem with respect to frequency goals and show that it is decidable in polynomial-time.
Finally, we provide some concluding remarks in Section~\ref{sec:conclusion}.



\section{Preliminaries} 
\label{sec:defs}

Let $X$ be a set and $i$ be a positive integer,
by $X^i$ we denote the cartesian product of $X$ with itself $i$ times and
by $X^*$ (resp., $X^\omega$) the set of finite (resp., infinite) sequences 
of elements of $X$.
By $\nats$, $\ints$, $\rats$, and $\reals$, we respectively denote
the set of non-negative integers, relative integers, rational, and real numbers.

For a positive integer $k$, let $\intv{k} = \{ 1, \ldots, k \}$.
A \emph{$k$-colored arena} is
a tuple $A = (V_0, V_1, \vini, E)$, where $V_0$ and $V_1$ are a partition
of a finite set $V$ of \emph{nodes}, $\vini$ is the \emph{initial node},
and $E \subseteq V \times \Sk \times V$ 
is a set of \emph{colored edges} such that for each node $v\in V$ 
there is at least one edge exiting from $v$. 
An edge $(u,a,v)$ is said to be \emph{colored} with $\acolor$.
In the following, we also simply call a $k$-colored arena an \emph{arena},
when $k$ is clear from the context.
For a node $v\in V$, we call $\efrom{v} = \{(v,a,w)\in E\}$ the set of edges
exiting from $v$, and $\eto{v} = \{(w,a,v)\in E\}$ the set of edges entering $v$.
For a color $a\in \intv{k}$, we call $E(a) = \{(v,a,w)\in E\}$
the set of edges colored with $a$.

A \emph{finite path} $\pat$ is a finite sequence of
edges $\{ (v_i,a_i,v_{i+1}) \}_{i \in \{0, \ldots, n-1\}}$, 
and its \emph{length} $|\pat|$ is the number of the edges it contains.
We denote by $\pat(i)$ the $i$-th edge of $\pat$.
Sometimes, we write the path $\pat$ as $v_0 v_1 \ldots v_n$,
when the colors are unimportant.
%
An \emph{infinite path} is defined analogously.
%
%
For a finite or infinite path $\pat$ and an integer $i$,
we denote by $\pat^{\leq i}$ the \emph{prefix} of $\pat$ containing
$i$ edges. 
The \emph{color sequence} of a finite (resp. infinite) path 
$\pat = \{(v_i, c_i,v_{i+1})\}_{i \in \{0, \ldots, n-1\}}$ 
(resp. $\pat = \{(v_i, c_i,v_{i+1})\}_{i\in \nats}$) on the arena $A$ 
is the sequence $\{c_i\}_{i \in \{0, \ldots, n-1\}}$
(resp.  $\{c_i\}_{i\in \nats}$)
of the colors of the edges of $\pat$.
When the meaning is clear from the context, we identify a path and its color sequence.
%
For all color sequences $x\in [k]^*$ and for all colors $\acolor,\bcolor\in [k]$,
we denote by $|x|_\acolor$ the number of edges colored with $\acolor$ in $x$,
and we set $\diff_{\acolor,\bcolor}(x) =  |x|_\acolor - |x|_\bcolor$.
The \emph{color difference matrix} of $x$, denoted $\diff(x)$,
is the $k \times k$ matrix whose generic element is 
$\diff(x)_{\acolor,\bcolor} = \diff_{\acolor,\bcolor}(x)$.

A \emph{$k$-colored game} is a pair $G=(A, W)$,
where $A=(V_0,V_1, \vini, E)$ is a $k$-colored arena
and $W\subseteq [k]^{\omega}$ is a set of color sequences
called \emph{goal}.
We assume that the game is played by two players, 
referred to as player $0$ and player $1$.
The players construct a path starting at $\vini$ on the arena $A$,
such a path is called \emph{play}.
Once the partial play reaches a node $v\in V_0$,
player $0$ chooses an edge exiting from $v$ and extends the play with
this edge; once the partial play reaches a node $v\in V_1$,
player $1$ makes a similar choice.
Player $0$'s aim is to make the play have color sequence in $W$, while player $1$'s aim is the opposite.
For $h\in \{0,1\}$, let $E_h = \{ (v,c,w) \in E \mid w\in V_{h}\}$ be the set of
edges ending into nodes of player $h$. Let $\varepsilon$ be the empty word,
a \emph{strategy} for player $h$ is a function
$\sigma_h : \varepsilon \cup (E^* E_h) \to E$ such that,
if $\sigma_h(e_0 \ldots e_n) = e_{n+1}$, then the destination of $e_n$
is the source of $e_{n+1}$, and if $\sigma_h(\varepsilon) = e$,
then the source of $e$ is $\vini$.
Intuitively, $\sigma_h$ fixes the choices of player $h$ for the entire game,
based on the previous choices of both players.
The value $\sigma_h(\varepsilon)$ is used to choose the first edge in the game.
A strategy $\sigma_h$ is \emph{memoryless} iff its choices
depend only on the last node of the play, i.e., for all plays $\rho$ and $\rho'$
with the same last node, it holds that $\sigma_h(\rho)=\sigma_h(\rho')$.
An infinite play $\{ e_i \}_{i \in \nats} \in E^{\omega}$
is \emph{consistent} with a strategy $\sigma_h$ iff
\emph{(i)} if $\vini \in V_h$ then $e_0 = \sigma_h(\varepsilon)$, and
\emph{(ii)} for all $i \in \nats$, if $e_i \in E_{h}$ then
$e_{i+1} = \sigma_h(e_0 \ldots e_i)$.
Note that,
given two strategies, $\sigma$ for player $0$ and $\tau$ for player $1$, there
exists only one play consistent with both of them.
We call such a play $P_{G}(\sigma, \tau)$.
A strategy for player $h$ is said \emph{winning} iff
all plays consistent with that strategy are winning for player $h$.
A game is said \emph{determined} iff one of the two players has a winning strategy.

Now we recall some definitions and results developed in \cite{Kop06}.
A goal $W\subseteq [k]^{\omega}$ is said to be \emph{prefix independent} iff for all
color sequences $x \in [k]^{\omega}$, and for all $z\in [k]^*$, we have
$x\in W$ iff $zx\in W$.
For two color sequences $x,y\in [k]^{\omega}$, the \emph{shuffle} of $x$ and
$y$, denoted by $\shuffle{x}{y}$ is the language of all the words $z_1 z_2 z_3
\ldots \in [k]^{\omega}$, such that $z_1 z_3 \ldots z_{2h+1} \ldots = x$  and
$z_2 z_4 \ldots z_{2h} \ldots = y$, where $z_i\in [k]^*$ for all $i\in \nats$.
A goal $W$ is said to be \emph{convex} iff it is closed w.r.t. the shuffle operation,
i.e., for all words $x, y \in W$ and $\shuffle{x}{y} \subseteq W$.
%
\begin{theorem}{\cite{Kop06}}
\label{thm:cop}
Let $G=(A,W)$ be a $k$-colored game such that $W$ is prefix-independent 
and convex. Then, the game is determined. Moreover, if player $1$ has a winning
strategy, he has a memoryless winning strategy.
\end{theorem}
%

\begin{section}{Colored Games with Frequency Goals}
\label{sec:balgames}

Let $\rho$ be an infinite path,
the \emph{frequency} of a color $a\in [k]$ on $\rho$ is the limit
$f_a = \lim_{n \rightarrow +\infty} \frac{|\rho^{\leq n}|_a}{n}$.
If such a frequency exists for all colors, then
the \emph{color frequency vector} of $\rho$ is $(f_1,\ldots,f_k)$.
It is trivial to prove that $\sum_{a=1}^k f_a=1$.
An infinite path $\rho$ is \emph{balanced} iff the frequency of each color $a\in [k]$
on $\rho$ is $f_a=\frac{1}{k}$;
$\rho$ has the \emph{bounded difference property} (in short, \emph{is bounded}) iff
there exists a constant $C\in \nats$ such that for all colors $a,b\in [k]$
and for all $n\in \nats$, $\diff_{a,b}(\rho^{\leq n})\leq C$.


%

In the following, we study $k$-colored games having one
of the following goals.

\begin{enumerate}
\item The \emph{bounded} goal $W_{bn}$, containing all and only
      the bounded color sequences.

\item The \emph{balance} goal $W_{bl}$,
      containing all and only the balanced color sequences.

\item Let $f \in \reals^k$ be such that $\sum_{i=1}^k f_i =1$.
  The \emph{frequency-$f$} goal $W_f$, containing
  all and only the color sequences with color frequency vector $f$.
\end{enumerate}

It is trivial to prove that the bounded, balanced, and frequency-$f$
goals are prefix-independent,
i.e., they do not depend on any finite prefix of a sequence.
%
The following lemma states two basic properties of the above goals.
\begin{lemma}{\cite{BFMM09}} \label{lem-props}
The following properties hold:
\begin{enumerate}
\item \label{one} if a path has the bounded difference property, then it is balanced;
\item \label{two} a path $\rho$ is balanced if and only if for all $a \in [k-1]$,
it holds $\lim_{i \to +\infty} \frac{\diff_{a,k}(\rho^{\leq i})}{i} =0$.
\end{enumerate}
\end{lemma}
The following example shows that the converse of item~\ref{one}
of Lemma~\ref{lem-props} does not hold.
%
\begin{example}{\cite{BFMM09}} \label{ex:path}
For all $i>0$, let $\sigma_i = (1 \cdot 2)^{i}
\cdot 1 \cdot 3 \cdot (1 \cdot 3 \cdot 2 \cdot 3)^{i}  \cdot 1 \cdot 3 \cdot
3$.
Consider the infinite sequence $\sigma = \prod_{i = 1}^{\omega} \sigma_i$
obtained by a hypothetic $3$-colored arena.
On one hand, it is easy to see that for all $i>0$ it holds
$\diff_{3,1}(\sigma_i) = 1$.
Therefore, $\diff_{3,1}(\sigma_1\sigma_2 \ldots \sigma_n ) = n$,
and $\sigma$ is not a bounded difference path.

On the other hand, since the length of the first $n$ blocks is $\Theta(n^2)$
and the difference between any two colors is $\Theta(n)$,
in any prefix $\sigma^{\leq i}$
the difference between any two colors is in $\bigo(\sqrt{i})$.
According to item~\ref{two} of Lemma~\ref{lem-props}, $\sigma$ is balanced.
\end{example}

\subsection{A Scheduling Example}

Consider two jobs in a concurrent program, both having the structure
shown in Figure~\ref{fig:job}.
Notice that the jobs exhibit nondeterministic behavior,
due to the unknown (i.e., not explicitly modeled) branching condition
on line~1.

\begin{figure}[h]
\begin{minipage}[t]{0.50\textwidth}
\vspace{-4.82truecm}
\begin{verbatim}
   while (1) {
0:   lock();
1:   if (...) {
2:     action();
     } else {
3:     action();
4:     action();
     }
5:   unlock();
   }
\end{verbatim}
\caption{A job in a concurrent program.\label{fig:job}}
\end{minipage}
\hfill
\begin{minipage}[t]{0.50\textwidth}
    \begin{tikzpicture}[node distance=1.3cm, auto, bend angle=20, shorten >=2pt, shorten <=2pt]
      \node[p1] (0-0) at (0,0)     {0,0};
      \node[p2] (1-0) at (-1.5,-1) {1,0};
      \node[p2] (0-1) at (1.5,-1)  {0,1};
      \node[p2] (2-0) [below left  of=1-0] {2,0};
      \node[p2] (3-0) [below right of=1-0] {3,0};
      \node[p2] (5-0) [below of=2-0] {5,0};
      \node[p2] (4-0) [below of=3-0] {4,0};
      \node[p2] (0-2) [below right of=0-1] {0,2};
      \node[p2] (0-3) [below left  of=0-1] {0,3};
      \node[p2] (0-4) [below of=0-3] {0,4};
      \node[p2] (0-5) [below of=0-2] {0,5};

      \path[-stealth'] (0-0) edge (1-0)
                             edge (0-1)
                       (1-0) edge (2-0)
                             edge (3-0)
                       (2-0) edge node[auto] {0} (5-0)
                       (3-0) edge node[auto] {0} (4-0)
                       (4-0) edge node[auto] {0} (5-0)
                       (5-0) edge [bend left=110] (0-0)
                       (0-1) edge (0-2)
                             edge (0-3)
                       (0-2) edge node[auto]      {1} (0-5)
                       (0-3) edge node[auto]      {1} (0-4)
                       (0-4) edge node[auto,swap] {1} (0-5)
                       (0-5) edge [bend right=110] (0-0)
                       ;
    \end{tikzpicture}
\caption{The non-preemptive scheduling game corresponding
to two jobs of the type in Figure~\ref{fig:job}.\label{fig:scheduling}}
\end{minipage}
\end{figure}

Assume we want to synthesize a scheduler that ensures that the
``{\tt action}'' function is called with the same asymptotic frequency
by the two jobs. The scheduler can decide not to give the lock
to a job, but cannot pre-empt them.
To this aim, we can produce a game as in Figure~\ref{fig:scheduling}, where
nodes represent joint configurations of the two jobs.
The only node of player~0 is represented by a circle,
while the nodes of player~1 are represented by boxes.
Since we are only interested in counting the calls to the {\tt action} function,
we only color the edges representing such call.
Clearly, uncolored edges can be represented in our framework by a sequence of
two edges, each labeled by a different color.
The internal nondeterminism of the jobs is modeled by a move of player~1.
The only choice for player~0 (the scheduler) occurs in node
$0,0$, where both jobs are waiting on the {\tt lock} operation,
and the scheduler can choose whom to give the lock to.

It is easy to verify that the scheduler has a strategy enforcing
the bounded difference property (hence, the balance property as well):
When the game is in $0,0$, give the lock to the job that executed
the {\tt action} function less times so far.
According to this scheduling policy, the difference between the number of 0's
and the number of 1's along a play will always be at most 2,
regardless of the choices made by the internal nondeterminism of the jobs.
Notice that this strategy requires memory.
Using a similar strategy, player~0 can also win w.r.t. the frequency-$f$ goal,
for all (rational) frequency vectors $f$.

\begin{subsection}{Co-NP Membership}

In this section, we prove that the problem of deciding whether there exists a winning
strategy for player $0$ in the games addressed in the previous section is in \conp.

\begin{lemma}{} \label{lemma:concavity}
$W_{bn}$, $W_{bl}$, and $W_{f}$ are convex.
\end{lemma}

\begin{proof-noq}
Let $y,z \in [k]^{\omega}$ and $x\in \shuffle{y}{z}$.
We prove that if $y$ and $z$ are both
balanced (resp., bounded, or frequency-$f$), then so is $x$.
We have that
$x=x_1\ldots x_i \ldots$ where $y=x_1 x_3 \ldots x_{2k+1} \ldots$ and
$z=x_2 x_4 \ldots x_{2k} \ldots$.
Also, for all $n\in \nats$
there are two indexes $n_y, n_z$ such that $n = n_y + n_z$ and
$\diff_{a,b}(x^{\leq n}) = \diff_{a,b}(y^{\leq n_y}) + \diff_{a,b}(z^{\leq n_z})$,
for all $a,b \in [k]$.
We distinguish the following cases.

\begin{enumerate}
\item (\textit{bounded})
Since $y$ and $z$ are bounded, there exist two constants
$C_{y}, C_{z}\in \nats$ such that for all $a,b\in [k]$
and for all $n>0$, $|\diff_{a,b}(y^{\leq n})| < C_y$ and
$|\diff_{a,b}(z^{\leq n})| < C_z$.
%
%
Therefore, let $C_{x} = C_{y}+C_{z}$,
for all $a,b\in [k]$ and $n\in \nats$ we have
$|\diff_{a,b}(x^{\leq n})| \leq |\diff_{a,b}(y^{\leq n_y})| +
                               |\diff_{a,b}(z^{\leq n_y})| \leq C_{x}$.
Hence, the sequence $x$ is bounded.

%
%
%
%

\item (\emph{frequency-$f$})
Given that $y$ and $z$ have frequency $f$,
we have that, for all $a\in [k]$ and for all $\varepsilon > 0$,
there exists $h(\varepsilon)>0$
such that for all $n>h(\varepsilon)$, it holds that
$\big| \frac{|y^{\leq n}|_a}{n} - f_{a} \big| \leq \varepsilon$ and
$\big| \frac{|z^{\leq n}|_a}{n} - f_{a} \big| \leq \varepsilon$.
%
Hence, given $\varepsilon>0$, let $n>0$ be such that $n_y \geq h(\varepsilon/2)$
and $n_z \geq h(\varepsilon/2)$. Such $n$ exists, due to the definition of the shuffle
operation. For all $n' > n$ we have that:
\begin{align*}
\Big| \frac{|x^{\leq n'}|_a}{n'} - f_{a} \Big| &=
\Big| \frac{|y^{\leq n'_y}|_a + |z^{\leq n'_z}|_a - (n'_y+n'_z)
f_{a}}{n'_y + n'_z} \Big| \\
&\leq \Big| \frac{|y^{\leq n'_y}|_a - n'_y \cdot f_{a}}{n'_y+n'_z} \Big| +
      \Big| \frac{|z^{\leq n'_z}|_a - n'_z \cdot f_{a}}{n'_y+n'_z} \Big| \\
&\leq \Big| \frac{|y^{\leq n'_y}|_a}{n'_y} - f_{a} \Big| +
      \Big| \frac{|z^{\leq n'_z}|_a}{n'_z} - f_{a} \Big|
\leq \varepsilon.
\end{align*}
So, the color sequence $x$ has frequency vector $f$.

\item
(\textit{balanced})
Since the balance property is equivalent to the frequency-$f$ property
with $f_{a}$ equal to $1/k$ for all colors $a\in [k]$, the thesis holds. \qed
\end{enumerate}
\end{proof-noq}

\noindent
Now, we can apply Theorem~\ref{thm:cop} to our goals and obtain the following.

\begin{corollary}{} \label{lmm:memorylessStrategy}
Let $G$ be a $k$-colored game with balance, bounded, or frequency-$f$ goal.
Then, the game is determined. Moreover if player $1$ has a winning
strategy, he has a memoryless winning strategy.
\end{corollary}
The fact that memoryless strategies suffice for player $1$
easily leads to the following result.

\begin{lemma}{} \label{lmm:CoNPMembership}
Given a $k$-colored game with balanced, bounded, or frequency-$f$ goal, 
the problem asking whether there exists
a winning strategy for player $1$ is in NP,
the problem asking whether there exists a winning strategy for player $0$ is in Co-NP.
\end{lemma}

\begin{proof}
By Corollary~\ref{lmm:memorylessStrategy},
if player $1$ has a winning strategy, he has a memoryless one.
The number of memoryless strategies is finite and
each one of them can be represented in polynomial space in the size of the problem.
So, in polynomial time we can guess a memoryless strategy $\tau$,
and verify that it is a winning strategy, using the following algorithm.
We construct the subarena $A'$, obtained from $A$ by removing all
the edges of player~$1$ that are not used by $\tau$.
We have that $\tau$ is a winning strategy for player $1$ in $A$ iff
all the plays on $A'$ are winning for player $1$.
Thus, player $0$ is able to construct a balanced
(resp. bounded, frequency-$f$)
path iff there exists a balanced (resp. bounded, frequency-$f$)
path in the graph of $A'$ and this path is reachable from $\vini$.
So, we construct the subgraph $A''$ of $A'$,
obtained by removing all the nodes that are not reachable from $\vini$.
In order to check if there exists a balanced (resp. bounded)
path reachable from $\vini$, it is sufficient to apply
the known polynomial-time algorithm~\cite{BFMM09}.
For the frequency-$f$ goal, a suitable polynomial-time algorithm
is presented in Section~\ref{sec:graphs}.

This concludes the proof that the problem of asking
whether there exists a winning strategy for player $1$ is in NP.
Hence, the complementary problem asking whether there exists
a winning strategy for player $0$ is in Co-NP.
\end{proof}

\end{subsection}

\begin{subsection}{Co-NP Hardness}

\begin{lemma}{} \label{lmm:CoNPHardness}
Given a boolean formula $\psi$ in conjunctive normal form, there exists a $k$-colored arena $A$ such that the following are equivalent
(i) $\psi$ is a tautology, (ii) there exists a winning strategy for player $0$ in the
game $G=(A, W_{bl})$, and (iii) there exists a winning strategy for player $0$ in the
game $G=(A, W_{bn})$.
\end{lemma}

\begin{proof}
Let $n$ be the number of clauses of $\psi$ and $m$ be the number of its variables,
then we can write $\psi=\wedge_{i=1}^n \psi_i$,
where each $\psi_i$ is a disjunction of literals.
In the following we define $\psi(x)$ 
as the set of all clauses in which $x$ appears in positive form,
and $\psi(\neg{x})$ 
as the set of all clauses in which $x$ appears negated.

\begin{figure}[ht]
	\label{fig:reduction}
    \begin{center}
        $$\xymatrix{
            & v_{j,1} \ar@/^1pc/[r] \ar@{.>} @/_1pc/[r]_{1}
            & v_{j,2}  \ar@/^1pc/[r] \ar@{.>} @/_1pc/[r]_{2}
            & \ldots \ar@/^1pc/[r] \ar@{.>} @/_1pc/[r]_{n}
            & v_{j,n+1} \ar[dr] \\
            v_{j}
\ar[ur] \ar[dr]
            &
            &
            &
            &
            & v'_{j} \\
            & \neg{v}_{{j},1}  \ar@/^1pc/[r]
			\ar@{.>} @/_1pc/[r]_{1}
            & \neg{v}_{j,2}  \ar@/^1pc/[r]
			 \ar@{.>} @/_1pc/[r]_{2}
            & \ldots \ar@/^1pc/[r] \ar@{.>} @/_1pc/[r]_{n}
            & \neg{v}_{j,n+1}  \ar[ur] \\
                            }
        $$
        \caption{\label{fig:prof}
          The $j$-th subgraph $A_j$ of $A$.
		  The dotted edges from $v_{j,i}$ to $v_{j,i+1}$ is present iff
		  $\psi_i \in \psi(x_j)$, and analogously for the lower branch. }
    \end{center}
\end{figure}
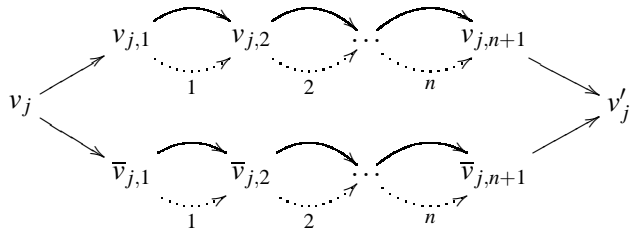

We construct the following $(n+1)$-colored arena $A=(V_0,V_1, \vini, E)$,
where the set of colors corresponds to the set of clauses of $\psi$ with
the added control color $n+1$.
%
The description of the arena $A$ makes use of
\emph{uncolored edges}, i.e., edges not labeled by any color. Clearly,
such an edge can be represented in our framework by a sequence of
$n+1$ edges, each labeled by a different color.
The arena $A$ is composed by $m$ subarenas $A_j$, one for each variable $x_j$. Every subarena $A_j$ has a starting node $v_j$, an ending node
$v'_j$ and two sequences of nodes: $\{v_{j, i}\}_{i=1}^{n}$,
$\{\neg{v}_{j, i}\}_{i=1}^{n}$ where every node is associated
with a clause.
There is an uncolored edge from $v_j$ to $v_{j, 1}$ and from
$v_j$ to $\neg{v}_{j, 1}$. Moreover, if we define $v_{j, n+1}=\neg{v}_{j, n+1}=v'_j$,
we have that for all $1\leq i \leq n$, \emph{(i)} there is an uncolored
edge from $v_{j, i}$ to $v_{j, i+1}$ and from $\neg{v}_{j, i}$ to $\neg{v}_{j, i+1}$,
\emph{(ii)} if $\psi_i\in \psi(x)$ then there is an $i$-colored edge from $v_{j,i}$ to $v_{j,i+1}$, and \emph{(iii)} if $\psi_i\in \psi(\neg{x})$ then there is an $i$-colored edge from $\neg{v}_{j,i}$ to $\neg{v}_{j, i+1}$.
We call the sequence $\{v_{j, i}\}_i$ the \emph{upper branch} of $A_j$ and
the sequence $\{\neg{v}_{j, i}\}_i$ the \emph{lower branch} of $A_j$.
The arena $A$ is constructed by connecting the subarenas $A_j$ as follows:
for all $1\leq j\leq m-1$ there is an uncolored edge from $v'_{j}$ to $v_{j+1}$ and
an $n+1$-colored edge from $v'_m$ to $v_1$.


%
The construction of $A$ is concluded by partitioning
the set of nodes as follows: $V_1=\{v_1, \ldots, v_m\}$ and $V_0=V-V_1$.
Intuitively, every subarena $A_j$ represents a truth choice for the variable $x_j$.
This choice is made by player $1$ with the aim to skip the passage through some clauses.
On the other hand, as soon as there is the chance, player $0$ tries
to pass through each clause once during a single loop,
in order to balance the clauses' colors with the control color $n+1$.
Let $G=(A, W_{bl})$ and $G'=(A, W_{bn})$,
we now show the correctness of the above construction.
In the following, we write $\unknown{v}_{j,i}$ to mean either $v_{j,i}$ or $\neg{v}_{j,i}$.

(If) If $\psi$ is a tautology, then the winning strategy for player $0$
in both games $G$ and $G'$ may be summarized as follows: as soon as there is a chance,
pass through an edge of color $\psi_i$;
then, do not pass through such an edge again, until we pass again through $v_1$.
Formally,
the strategy of player $0$ is the following: each time the play is in a node
$\unknown{v}_{j,i}$, player $0$ chooses to reach $\unknown{v}_{j,i+1}$ through
the $\psi_i$-colored edge iff $\psi_i$ does not appear in the least suffix
of the partial play starting with $v_1$.
%
We observe that during a single loop from $v_1$ to itself, a strategy of player $1$
is a truth-assignment to the variables of $\psi$: precisely for every subarena $A_j$,
player $1$ chooses to follow the upper branch iff $x_j$ is true.
Since $\psi$ is a tautology, any such assignment is a satisfiable assignment,
i.e., given such an assignment $a:\{x_1,\ldots,x_n\} \rightarrow \{T,F\}$,
for each clause $\psi_i$, there exists a variable $x$ such that $\psi_i$ is true
also due to the value $a(x)$.
This means that player $0$ can pass through a $\psi_i$-colored edge at least once
during a single loop, and thanks to his strategy, he will pass through such an edge exactly once.
Thus, during each loop, the uncolored edges are already perfectly balanced,
and the edges added by player $0$ are balanced thanks to the last $n+1$-colored edge.
Thus, during the infinite play, the color differences are always zero when the play is 
in node $v_1$. Since the loops from $v_1$ to itself have bounded length,
the color differences are bounded during the play.
Thus every infinite play consistent with the strategy is bounded and it is balanced too, because in \cite{BFMM09} we proved that a bounded path is balanced too.

(Only If). If $\psi$ is not a tautology, then there is a memoryless winning strategy
 for player $1$ on $G$ and on $G'$: player $1$ follows a truth assignment
of the variables of $\psi$ that does not satisfy $\psi$.
For such an assignment there is an unsatisfiable clause $\psi_i$.
So, during a loop from $v_1$ to itself, if player $1$ follows this strategy,
player $0$ cannot pass through any $\psi_i$-colored edge.
Thus, at the end of the loop the color difference between color $\psi_i$
and color $n+1$ is increased by one.
Every play $\pat$ is an infinite concatenation of simple loops from $v_1$ to itself.
Since those loops have maximum length $l\leq |E|$,
for all $j\in \nats$ we have $\diff_{i, n+1}(\pat^{\leq j}) \geq \frac{j}{l}$,
and thus $\lim_{j\rightarrow +\infty} \frac{\diff_{i, n+1}(\pat^{\leq j})}{j}\geq \frac{1}{l}$.
This means that every play consistent with said strategy of player~$1$ is not balanced,
and hence not bounded.
\end{proof}

\begin{theorem}{}
Given a $k$-colored game $G$ with balanced (resp., bounded, frequency-$f$) goal,
the problem asking whether there exists a winning strategy
for player $0$ is Co-NP-complete.
\end{theorem}

\begin{proof}
By Lemma~\ref{lmm:CoNPMembership} and Lemma~\ref{lmm:CoNPHardness},
we have that the problems for the balance and the bounded goal
are Co-NP-complete.
Since the bounded goal is a special case of frequency-$f$ goal (for $f_i = 1/k$),
we have that the frequency-$f$ problem is \conp-hard too.
Since by Lemma~\ref{lmm:CoNPMembership}
the problem for frequency-$f$ is in \conp, it is \conp-complete.
\end{proof}

This \conp-completeness result may be regarded as essentially negative.
In fact, the algorithm showing membership in NP, once converted
into a deterministic form, simply suggests to try each one of the
(exponentially many) memoryless strategies of player~1 in the game,
and solve a linear program to determine whether it is winning.
It remains to investigate the possibility of practically efficient
algorithms, arising, for instance, from the analysis of the specific
properties of the games of interest.

\end{subsection}

\end{section}

\section{The Frequency-$f$ Problem on Graphs} \label{sec:graphs}

In this section, we show that if player~$0$ controls all nodes in a frequency-$f$
game, the existence of a winning strategy can be determined in polynomial time,
by reducing the problem to the feasibility of a linear system of equations.
In the following, a \emph{$k$-colored graph} is an arena
whose nodes belong all to player $0$.
%
%
We employ an alternative, essentially equivalent formulation of the frequency-$f$ goal,
called \emph{color-limit-$\limit$} goal.
We define the \emph{color limit} of an infinite path $\rho$
as the matrix $\{ l_{i,j} \} \in\reals^{k \times k}$,
where $l_{i,j} = \lim_{n \rightarrow +\infty}
\frac{\diff_{i,j}(\rho^{\leq n})}{n}$.

\begin{lemma}{} \label{lmm:colorLimitEquivalence}
An infinite path $\rho$ has color limit $L\in \reals^{k\times k}$ iff
its color frequency vector $f$ exists and it is the unique solution
of the following system of $k^2+1$ linear equations:
for all $i,j\in [k]$, $f_i-f_j = l_{i,j}$; $\sum_{i=1}^k f_i =1$.
\end{lemma}

\begin{proof}
First, observe that the system of linear equations $f_i-f_j=l_{i,j}$ and
$\sum_{i=1}^k f_i =1$ contains $k$ independent rows in the coefficient matrix,
i.e., the rows associated with the equations $f_1-f_k=l_{1,k}$, $\ldots$,
$f_{k-1}-f_k=l_{k-1,k}$, and $\sum_{i=1}^k f_i=1$.
So, the system may have only one solution or no solutions at all.

[only if] If $\rho$ has color frequency vector $f\in \reals^k$, 
then for all $i\in [k]$, it holds that
$f_i=\lim_{n\rightarrow +\infty} \frac{ |\rho^{\leq n}|_i}{n}$. So,
for all $i,j\in [k]$, it holds that
$l_{i,j}= \lim_{n\rightarrow +\infty} \frac{|\rho^{\leq n}|_i
- |\rho^{\leq n}|_j}{n}=f_i-f_j$.

[if] If $\rho$ has color limit $L$ then, for all $i,j\in [k]$, it holds that
$l_{i,j}= \lim_{n\rightarrow +\infty} \frac{\diff_{i,j}(\rho^{\leq n})}{n}$.
We show that \emph{(i)}
$\lim_{j \rightarrow +\infty} |\pat^{\leq j}|_{k}/j = l \stackrel{\Delta}{=}
\frac{1-\sum_{a\in [k-1]} l_{a,k}}{k}$;  and \emph{(ii)} for all $a\in[k-1]$,
the sequence $\{ |\pat^{\leq j}|_{a}/j\}_j$ converges to $l_{a,k}+l$.
%
First we show $(i)$.
Assume by contradiction that the sequence is not convergent
to $l$, then we have $\exists \varepsilon >0
\qdot \forall m\in \nats \qdot \exists n_{m}\geq m \qdot\; \allowbreak
\Big( \frac{|\pat^{\leq n_{m}}|_{k}}{n_{m}}>l+\varepsilon \;\text{ or }\;
\frac{|\pat^{\leq n_{m}}|_{k}}{n_{m}}<l-\varepsilon \Big). $
The points $\{n_{m}\}_m$ form a sequence, from which
we can extract two subsequences $\{n_{m_i}\}_i$, given by all the points
such that $|\pat^{\leq n_{m_i}}|_{k}/n_{m_i}>l+\varepsilon$, and
$\{n_{m'_i}\}_i$, given by all the points such that
$|\pat^{\leq n_{m'_i}}|_{k}/n_{m'_i}<l-\varepsilon$.
At least one of the two subsequences is infinite.
%
%
%
Assume w.l.o.g. that $\{n_{m_i}\}_i$ is infinite.
        Then, $\sum_{a=1}^{k-1} l_{a,k} =
        \sum_{a=1}^{k} \big( \frac{|\pat^{\leq n_{m_i}}|_{a}}{n_{m_i}}-l \big) >
        \sum_{a=1}^{k-1} \big( \frac{|\pat^{\leq n_{m_i}}|_{a}}{n_{m_i}}-l \big)
				+\varepsilon$. 
        In other words, 
        $\sum_{a=1}^{k-1} \big( \frac{|\pat^{\leq n_{m_i}}|_{a}}{n_{m_i}}-l \big) <
         \sum_{a=1}^{k-1} l_{a,k} - \varepsilon$.
        So, for all $i\in \nats$ there is a color
	$a\in [k-1]$ such that $\frac{|\pat^{\leq n_{m_i}}|_{a}}{n_{m_i}} -l \leq
	l_{a,k}-\frac{\varepsilon}{k-1}$.
	%
	%
	Then, there is a color $a\in [k-1]$ and a subsequence $\{n_{m^a_i}\}_i$
	of $\{n_{m_i}\}_i $ such that for all $i\in \nats$ we have that
	$\frac{|\pat^{\leq n_{m^a_i}}|_{a}}{n_{m^a_i}} <
	l_{a,k}+l-\frac{\varepsilon}{k-1}$.
	Moreover, for all $i\in \nats$ 
	we have
	$\frac{\diff_{a,k}\big(\pat^{\leq n_{m^a_i}}\big)}{n_{m^a_i}}$ =
        $\frac{ \big|\pat^{\leq n_{m^a_i}}\big|_{a} - \big|\pat^{\leq n_{m^a_i}}\big|_{k}}{n_{m^a_i}} \leq
	   \big( l_{a,k}+l- \frac{1}{k-1}\varepsilon \big) - (l+\varepsilon)=
	l_{a,k} - \frac{k}{k-1}\varepsilon$.
	%
	%
	Therefore, the sequence
	$\{ \diff_{a,k}(\pat^{\leq n_{m^a_i}})/n_{m^a_i} \}_i$
	does not converge to $l_{a,k}$,
	so does not the sequence $\{ \diff_{a,k}(\pat^{\leq j})/j\}_j$,
	since the first is a subsequence of the latter.
So, by contradiction, we have proved $(i)$.
%

Now, we show $(ii)$.
Assume by contradiction that $\{|\pat^{\leq j}|_{a}/j\}_j$
does not converge to $l+l_{a,k}$, for a certain $a \in\intv{k-1}$. Then, we have
	\begin{equation} \label{eq:01}
	\exists \varepsilon >0
	\qdot \forall m\in \nats \qdot \exists n_m\geq m \qdot
	\bigg( \frac{|\pat^{\leq n_m}|_{a}}{n_m}>l+l_{a,k} + \varepsilon \;\text{ or
}\;
	\frac{|\pat^{\leq n_m}|_{a}}{n_m}<l+l_{a,k}-\varepsilon \bigg).
	\end{equation}
Let $\myeps$ be a witness for~\eqref{eq:01}.
By~$(i)$, there is $\bar{n}\in \nats$ such that for all $n\geq\bar{n}$, we have
$l-\myeps/2 < |\pat^{\leq n}|_{k}/n < l+\myeps/2$.
%
%
So, for all $m \geq \bar{n}$, there is $n_m\geq m$ such that 
either \emph{(a)} $\frac{|\pat^{\leq n_m}|_{a}}{n_m} > l+l_{a,k} + \myeps$
or     \emph{(b)} $\frac{|\pat^{\leq n_m}|_{a}}{n_m} < l+l_{a,k}-\myeps$,
depending on which disjunction in \eqref{eq:01} holds.
Assuming that \emph{(a)} occurs for infinitely many $n_m$,
for all $m \geq\bar{n}$ there is $n_m\geq m$ such that
\begin{align*}
\frac{\diff_{a,k}(\pat^{\leq n_m})}{n_m} &=
\frac{|\pat^{\leq n_m}|_{a} - |\pat^{\leq n}|_{k}}{n_m} >
l + l_{a,k} + \myeps - \Big(l+ \frac{\myeps}{2}\Big) = l_{a,k}+\frac{\myeps}{2}.
\end{align*}
Thus, we have that $\{ \diff_{a,k}(\pat^{\leq n})/n \}_n$ does not
converge to $l_{a,k}$, which is a contradiction.
\end{proof}

Thus, the frequency-$f$ problem is equivalent to the problem asking whether
there exists a path $\rho$ with color limit $L$,
where $l_{i,j}=f_i-f_j$, for all $i,j\in [k]$.



We reduce the color-limit-$L$ problem to the feasibility of a system
of linear equations, using a technique similar to the one
we used to solve the balance problem on graphs~\cite{BFMM09}.
%
Due to their technicality, the proofs are postponed to Section~\ref{sec:proofs}.

\begin{definition}{}
  \label{def:balancedFeasabilitySet}
  Let $A$ be a $k$-colored graph, and $L\in \reals^{k\times k}$
  a square matrix.
  We call \emph{color-limit-$L$ system} for $A$
  the following system of equations
  on the set of variables $\{ x_e \mid e \in E\}$.
$$
\begin{array}{lrl}
\text{1. for all } v \in V  &\sum_{e\in \eto{v}} x_e &= \sum_{e\in \efrom{v}} x_e \\
\text{2. for all } a,b \in [k]\quad &\sum_{e\in E(a)} x_e -\sum_{e \in E(b)} x_e&
= l_{a,b} \sum_{e\in E} x_e \\
\text{3. for all } e\in E        &x_e &\geq 0 \\
\text{4.}          &\sum_{e\in E} x_e&> 0.
\end{array}
$$
\end{definition}
Let $m=|E|$ and $n=|V|$, the color-limit-$L$ system has $m$ variables
and $m+n+k^2+1$ constraints.
It helps to think of each variable $x_e$ as a load associated
to the edge $e\in E$, and of each constraint as having the following
meaning.

\begin{enumerate}
\item For each node, the entering load is equal to the
      exiting load.

\item For all colors $a,b\in [k]$, the difference between the loads
      on the edges colored by $a$ and by $b$ is equal to
      $l_{a,b}$ times the whole load.

\item Every load is non-negative.
\item The total load is positive.
\end{enumerate}
The following lemma states the reduction from the color limit-$L$ problem
to the feasibility of the system.
\begin{lemma}{} \label{lmm:colorLimitInPStep1}
In a graph $A$, there exists an infinite path with color limit $L$
iff the color-limit-$\limit$ system for $A$ is feasible.
\end{lemma}





Since the feasibility problem for a system of linear equations is solvable
in polynomial time in the size of the system
(number of constraints and size of the coefficients) \cite{NW88},
we obtain the following.

\begin{theorem}{} 
  The color-limit-$L$ problem is in PTIME.
\end{theorem}

As we show later in Lemmas \ref{lmm:1H} and \ref{lmm:colorLimitInPStepA},
it is possible to construct in polinomial time, from a solution of the linear
system, a representation of a path in the graph satisfying the frequency-$f$
constraint.

\def\loops{\mathcal{L}}

\subsection{Proof of Lemma~\ref{lmm:colorLimitInPStep1}} \label{sec:proofs}

We first need some additional notations and preliminary lemmas.
%
%
Let $A=\{(A_1,w_1),\ldots,(A_m,w_m)\} \subseteq \ints^{d\times d} \times \nats$ 
be a finite set of $m$ pairs (integer matrix, respective weight),
we call \emph{natural linear combination} (in short, \emph{n.l.c.})
of the elements of $A$ any matrix $D = \sum_{i=1}^m c_i A_i$, 
where each $c_i$ is a non-negative integer,
and at least one $c_i$ is strictly positive.
Moreover, we define the \emph{weight} of $D$ as $n_D = \sum_{i=1}^m c_i w_i$ and
the \emph{ratio} of $D$ as $\frac{D}{n_D}$.

Intuitively, we introduce this machinery to 
express properties of sets of simple loops in a colored graph.
Each simple loop $\rho$ in the set induces a (matrix, weight) pair,
where the $(i,j)$ element of the matrix contains the difference
between the occurrences of color $i$ and color $j$ in $\rho$,
and the weight is the length of $\rho$.
Given a set of loops, the integer coefficients of an n.l.c. $D$ represent
the number of times that each of the loops must be taken in order
to build some path of interest.
The weight of $D$ is simply the total length of the obtained path
and the ratio of $D$ is its color difference matrix, divided by the
length of the path.
Accordingly, we say that a matrix is an n.l.c. of a set of loops $\loops$
when it is an n.l.c. of the set 
$A=\{ (\diff(\sigma), |\sigma|) \mid \sigma\in\loops \}$. 
In the following, by $M^T$ we denote the transpose of the
matrix $M$ and by $M_{i,j}$ we denote the element of $M$ at its $i$-th row and 
$j$-th column. 
We say that a set of loops is \emph{connected} if the loops
belong to the same strongly connected component, or, 
equivalently, if they are pairwise mutually reachable. 

\begin{lemma}{} \label{lmm:1E}
  Let $L\in \rats^{d \times d}$, and $A \subset \ints^{d\times d} \times
  \nats$ be a finite set 
  such that no n.l.c.\ of $A$ has ratio $L$.
  Let $\{ (B_n, u_n) \}_{n}$ be an infinite
  sequence of elements of $A$,
  $S_{n}= \sum_{l = 0}^{n} B_l$ be the partial sum,
  and $U_n=\sum_{l= 0}^n u_l$ be the partial sum of the weights.
  Then, there exist two indexes $i,j\in [d]$
  such that $\lim_{n \rightarrow +\infty} \frac{S_{n,i,j}}{U_n} \neq L_{i,j}$.
\end{lemma}

\begin{proof}
  Let $A = \{ (A_1,w_1),\ldots, (A_m, w_m)\}$ and
  $f : \reals^{m} \mapsto \reals_{+}$
  be the function $f(c_{1}, \ldots, c_{m}) \allowbreak =
  \max_{1 \leq i,j \leq d} \big\{ \big| \frac{\sum_{n = 1}^{m}
    c_{n} A_{n,i,j}}{\sum_{n = 1}^{m} c_{n} w_n}- L_{i,j} \big| \big\}$.
  First, note that $f$ is a continuous function,
  since it is the maximum of continuous functions.
  Let now $K \subset \reals^{m}$ be the set $\{ (c_{1}, \ldots, c_{m}) \in
  [0,1]^{m} \mid \sum_{i = 1}^{m} c_{i} = 1 \}$.
  Note that $\zeroMatrix \not\in K$ and that $K$ is compact, since
  it is a finite dimensional space defined by a linear equation.
  Hence, by Weierstrass theorem, $f$ admits a minimum
  $M = \min_{x\in K} \{ f(x) \}$ on $K$.
  Since, by hypothesis, there is no n.l.c. of $A$ with ratio $L$, 
  $M$ must be strictly positive.
  Indeed, if by contradiction $M = 0$, there should be a non-zero vector
  $(c_{1}, \ldots, c_{m}) \in K$ such that for all $i,j\in [d]$,
  \begin{equation} \label{eq:internal}  
     \sum_{n = 1}^{m} c_{n} A_{n,i,j} - L_{i,j} \sum_{n = 1}^{m} c_{n} w_n=M=0.
  \end{equation}
  Since~\eqref{eq:internal} is a homogeneous linear equation with
  rational coefficients and since it has a non-negative solution,
  it also has a non-negative \emph{integer} solution with
  at least one positive component. 
  This solution induces a n.l.c. of $A$ with ratio $L$,
  contradicting the hypothesis on $A$.
  
  Now, consider the sequence $\{(B_n, u_n) \}_{n}$,
  its partial sums $S_{n} = \sum_{l = 0}^{n} B_l$, and
  its weight partial sum $U_n = \sum_{l = 0}^n u_l$.
  Moreover, let $\delta_{l,n}$ be the number of occurrences
  of $(A_l,w_l)$ in the sequence up to position $n$ 
  and let $c_{l,n} = \delta_{l,n} / n$.
  Then $S_n = \sum_{l = 1}^{m} \delta_{l,n}
  \cdot A_l = n \cdot \sum_{l = 1}^{m} c_{l,n} \cdot A_l$
  and 
  $U_n = \sum_{l = 1}^{m} \delta_{l,n}
  \cdot w_l = n \cdot \sum_{l = 1}^{m} c_{l,n} \cdot w_l$.
  Since we have $\sum_{l = 1}^{m} \delta_{l,n} = n$ for all $n \in
  \nats$, it is obvious that $(c_{1,n}, \ldots, c_{m,n}) \in K$.
  
  Let now $Z_{n}\in \reals^{d \times d}$ be the matrix 
  defined by $Z_{n,i,j}=\left| \frac{\sum_{l = 1}^{m}
    c_{l,n} A_{l,i,j}}{\sum_{l = 1}^{m} c_{l,n} w_l}- L_{i,j} \right|$.
  Since there is no n.l.c. of $A$ with ratio $L$, it holds that for
  all $n \in \nats$ there exists a non-zero element in $Z_{n}$.
  Let $\{ (i_n, j_{n}) \}_{n}$ be an index sequence such that
  $Z_{n,i_n,j_n} =
  \max_{1 \leq i,j \leq d} \{Z_{n,i,j} \} > 0$.
  Since the sequence $\{ (i_{n}, j_{n}) \}_{n}$ can assume at most $d^2$
  different values, there exists a pair $(i^*, j^*)$ that occurs infinitely often in it.
  Let $\{ h_{t} \}_{t}$ be the index sequence such that
  $(i_{h_{t}}, j_{h_t}) = (i^*, j^*)$ and
  there is no $t' \in ] h_{t}, h_{t+1} [$ with $(i_{t'}, j_{t'}) = (i^*, j^*)$.
  Then, consider the subsequence
  $\{Z_{h_{t}, i^*, j^*}\}_{t}$ of $\{ Z_{n, i^*, j^*} \}_{n}$.
  We obtain that $\lim_{t \rightarrow +\infty} Z_{h_{t}, i^*, j^*} \geq M > 0$
  and consequently that $\lim_{n \rightarrow +\infty} Z_{n, i^*, j^*} \neq 0$, 
  whenever these limits exist.
  In conclusion, $\lim_{n \rightarrow +\infty} \frac{\sum_{l = 1}^{m}
	  c_{l,n} A_{l,i,j}}{\sum_{l = 1}^{m} c_{c,l} w_l} = 
          \lim_{n \rightarrow +\infty} \frac{S_{n,i,j}}{U_n} \neq L_{i,j}$.
\end{proof}

The next lemma uses the concept of \emph{quasi-segmentation}.
Intuitively, the quasi-segmentation of a path is a partition of the path 
in a sequence of simple loops and in a residual simple path.
For a finite path $\rho$, we define the quasi-segmentation and the 
\emph{rest} recursively on the length (i.e. the number of edges) 
of $\rho$ as follows.
The quasi-segmentation is always a finite sequence of loops, 
and the rest is a simple path ending with the last node of $\rho$.
If $\rho$ has length $1$ and it is not a loop, then the quasi-segmentation is
the empty sequence and the rest is $\rho$ itself.
If $\rho$ has length $1$ and it is a loop, then the quasi-segmentation is
$\rho$ itself and the rest is the last node of $\rho$.
If $\rho$ has size $n$,
let $\rho'=\rho^{\leq n-1}$, let $\sigma_1,\ldots,\sigma_n$ be the quasi segmentation
of $\rho'$ and $r$ be its rest. 
Consider the path $r'$ obtained by extending $r$ with the last edge of $\rho$
(this can be done because the last node of $r$ is the last node of $\rho'$).
If $r'$ does not contain a loop, then the quasi-segmentation of $\rho$ is
$\sigma_1,\ldots,\sigma_n$ and the rest is $r'$.
If $r'$ contains a loop $\sigma$, this loop is due to the last added edge, 
i.e., $r' = r'' \sigma$. 
In this case the quasi-segmentation of $\rho$ is $\sigma_1,\ldots, \sigma_n,\sigma$.
Moreover if $r''$ is non-empty the rest is $r''$ 
(note that $r''$ ends with the last node of $\rho$ because $r''$ ends with the first node of $\sigma$ which is the last node of $\sigma$, too).
Otherwise, if $r''$ is empty the rest is the last node of $\rho$.
%
%
The quasi-segmentation of an infinite path $\rho$ is the infinite sequence
of loops given by the limit of the quasi-segmentation of $\rho^{\leq n}$, for
$n\rightarrow +\infty$. An infinite path has no rest.

\begin{lemma}{}	\label{lmm:colorLimitSet}
		Let $G$ be a $k$-colored graph and $\rho$ be an infinite path in $G$
		with color limit $L\in \reals^{k\times k}$, then there exists a connected
		set of simple loops having an n.l.c. with ratio $L$.
\end{lemma}

\begin{proof}
		Since $\pat$ is an infinite path over a finite set of nodes,
		there exists a non-empty set $V'$ of nodes through which the path
		passes an infinite number of times.
		Then, there exists a constant $m$ such that, for all $n\geq m$,
		it holds that $\pat(n)\in V'$.
		The path $\pi \stackrel{\Delta}{=} \pat^{\geq m}$
		has color limit $L$, since the color-limit property is prefix independent.
		Let $\{\sigma_i\}_i$ be the quasi-segmentation of $\pi$ and,
		for all $i\in\nats$, let $h(i)$ be the index in $\pi$ of the node in which
		$\sigma_i$ closes itself.
		So, each time a simple loop closes at step $h(n)$, $\pi^{\leq h(n)}$ is
		composed by the $n+1$ simple loops $\sigma_0,\ldots, \sigma_n$, closed so
		far plus the rest $r_n$.
%
		Then, let $\loops$ be the set of
		all simple loops in the graph $G$, and let
		$A = \{(\diff(\sigma), |\sigma|) \mid \sigma \in \loops\}$.
		For all $i,j\in [k]$, let $\diff_{n,i,j} = \diff_{i,j}(\pi^{\leq n})$.
		Since $\pi$ has color limit $L$, we have
		$\lim_{n\rightarrow + \infty} \frac{\diff_{h(n),i,j}}{h(n)} = L_{i,j}$,
		for all $i,j\in [k]$.
		We observe that $\{\diff(\sigma_n)\}_n$ is a sequence of elements of $A$.
		Let $S_{n}=\sum_{i=1}^n \diff(\sigma_i)$ be the partial sum and
		$W_n=\sum_{i=1}^n |\sigma_i|$ be the partial sum of the lengths.
		So, for all $i,j \in [k]$, we have $\diff_{h(n),i,j}=
		\diff_{i,j}(r_n) + \sum_{q=1}^n \diff_{i,j}(\sigma_q)$.
		
		Since the rest is a simple path, it has length at most $|V'|$, 
		and we have $S_{n,i,j} - |V'| \leq \diff_{h(n),i,j} \leq
		S_{n,i,j} + |V'|$.
		Hence, $|S_{n,i,j}-\diff_{h(n),i,j}|\leq |V'|$ and $\diff_{h(n),i,j}-|V'|\leq
		S_{n,i,j}\leq \diff_{h(n),i,j}+|V'|$.
		Moreover, $h(n) = |r_n| + \sum_{q=1}^n |\sigma_q|$,
		thus $W_n-|V'| \leq h(n) \leq W_n + |V'|$,
		so we have $h(n) - |V'| \leq W_n  \leq h(n)+ |V'|$.
		For all $i,j\in [k]$, since
		$\lim_{n\rightarrow +\infty} \frac{\diff_{h(n),i,j}}{h(n)}
		=L_{i,j}$, then $\lim_{n\rightarrow +\infty}
		\frac{\diff_{h(n),i,j}+|V'|}{h(n)-|V'|}=L_{i,j}$ and
		$\lim_{n\rightarrow +\infty}
		\frac{\diff_{h(n),i,j}-|V'|}{h(n)+|V'|}=L_{i,j}$.
		Since for all $n\in\nats$ such that $h(n)>|V'|$ we have
		$\frac{\diff_{h(n),i,j}-|V'|}{h(n)+|V'|}
		\leq \frac{S_{n,i,j}}{W_n}\leq
		\frac{\diff_{h(n),i,j}+|V'|}{h(n)-|V'|}$,
		we have $\lim_{n\rightarrow +\infty} \frac{S_{n,i,j}}{W_n}=L_{i,j}$.
		%
		By Lemma~\ref{lmm:1E}, $A$ has an n.l.c. $D$ with ratio $L$.
%
		Then, the simple loops of $\loops$
		which occur with a positive coefficient in $D$ are connected,
		because they are extracted from the same path $\pi$,
		and have an n.l.c. with ratio $L$.
\end{proof}

In Lemma~\ref{lmm:1H}, we show how to construct a path with a given color
limit from a connected set of simple loops. 
The next lemma is needed as an auxiliary result.
Informally, it allows us to state that if on the path we find some points,
whose distance grows quadratically, while the color differences grow linearly
along those points, then the color limit exists and depends on the rate of
this growth.

\begin{lemma}{}	\label{lmm:1G}
		Let $\{ a_{n} \}_{n}$ be a sequence of integers, $c,c',c'',k \in \ints$,
		and $\{ x_{i}\}_{i}$ be an index sequence such that for all
		$i \in \nats$ it holds that \emph{(i)}
		$x_{1} = 1$, \emph{(ii)} $x_{i+1} \geq x_{i}$, \emph{(iii)} $x_{i+2} -
		x_{i+1} = x_{i+1} - x_{i} + k$, and \emph{(iv)}
		$c +c' \cdot i + \min \{ a_{n} \mid n \in
		[ x_{i}, x_{i+1} [\} \leq \min \{ a_{n} \mid n
		\in [x_{i+1}, x_{i+2} [ \} $ and $ \max \{ a_{n} \mid n
		\in [x_{i+1}, x_{i+2} [ \} \leq c'' +c' \cdot i + \max \{a_{n} \mid n \in
		[ x_{i}, x_{i+1} [\}$.
		Then, $\lim_{n \rightarrow +\infty} \frac{a_{n}}{n} = \frac{c'}{k}$.
\end{lemma}

\begin{proof}
		Let $\{ b_{n} \}_{n}$ and $\{ m_{i} \}_{i}$ be two sequences such that
		$b_{n} = a_{m_{i}} = \max \{ a_{n} \mid n \in [ x_{i}, x_{i+1} [ \}$,
		for all $n \in [ x_{i}, x_{i+1} [$ and $i \in \nats$.
		Obviously $a_{n} \leq b_{n}$.
		Moreover, let $\{ k_{n} \}_{n}$ be a sequence for which it holds that $n \in
		[ x_{k_{n}}, x_{k_{n}+1} [$.
		Then, by construction we can observe that $k_{1} = 1$, $b_{n} =
		|a_{m_{k_{n}}}|$, and $|a_{m_{i}}| \leq c + c'\cdot i + a_{m_{i-1}}
		\leq \ldots \leq (i-1) \cdot c + c'(\sum_{j=2}^i j) + a_{m_{1}}$,
		so it holds that $b_{n} \leq (k_{n}-1)\cdot c+\frac{1}{2}c'(k_n^2-k_n-1)
		+b_{1}$. Consider now the fraction $\frac{b_{n}}{n}$.
		Since $n \in [ x_{k_{n}}, x_{k_{n}+1} [$, we have $\frac{b_{n}}{n} \leq
		\frac{(k_{n}-1) \cdot c +\frac{1}{2}c'(k_n^2-k_n-1)+ b_{1}}{x_{k_{n}+1}}
		\leq \frac{(k_{n}-1) \cdot c + \frac{1}{2}c'(k_n^2-k_n-1) + b_{1}}{1 +
		\sum_{i = 1}^{k_{n}} (x_{i+1} - x_{i})}$.
		By the hypothesis on the sequence $\{ x_{i} \}_{i}$, there is a
		constant $k_0$ such that
		$x_{i+1} - x_{i} = (x_2-x_1) + \sum_{j=2}^i k = k_0 + k(i-1)$,
		so we have $\frac{b_{n}}{n} \leq
		\frac{(k_{n}-1) \cdot c + \frac{1}{2}c'(k_n^2-k_n) + b_{1}}
		{1 + \sum_{i = 1}^{k_{n}} (k_0+k(i-1))}= \allowbreak
		\frac{(k_{n}-1) \cdot c + \frac{1}{2}c'(k_n^2-k_n) + b_{1}}
		{1 + k_0i+\frac{1}{2}k(k_n^2+k_n-1)}$.

		Let $\{ b'_{n} \}_{n}$ and $\{ m'_{i} \}_{i}$ be two sequences such that
		$b'_{n} = a_{m'_{i}} =\min\{ a_{n} \mid n \in [ x_{i}, x_{i+1} [ \}$,
		for all $n \in [ x_{i}, x_{i+1} [$ and $i \in \nats$.
%
		Dually we can prove that $\frac{b'_{n}}{n} \geq
		\frac{(k_{n}-1) \cdot c'' + \frac{1}{2}c'(k_n^2-k_n) + b_{1}}
		{1 + k_0i+\frac{1}{2}k(k_n^2+k_n-1)}$.

		So, for all $n\in \nats$ $\frac{(k_{n}-1) \cdot c'' +
		\frac{1}{2}c'(k_n^2-k_n) + b_{1}}{1 + k_0i+\frac{1}{2}k(k_n^2+k_n-1)}\leq
		\frac{b'_n}{n}\leq \frac{a_n}{n} \leq \frac{b_n}{n} \leq
		\frac{(k_{n}-1) \cdot c' + \frac{1}{2}c'(k_n^2-k_n) + b_{1}}
		{1 + k_0i+\frac{1}{2}k(k_n^2+k_n-1)}$.
		Since the extremes converge to $\frac{c'}{k}$ as $n$ goes to
		infinity, we have
		$\lim_{n \rightarrow +\infty} \frac{a_{n}}{n} = \frac{c'}{k}$.
\end{proof}

\begin{lemma}{} \label{lmm:1H}
		If a $k$-colored graph $G$ contains a set of connected
		simple loops having an n.l.c. of ratio $L$,
		then there exists in $G$ an infinite path $\rho$ with color
		limit $L$.
\end{lemma}

\def\amax{\mathit{AM}}
\def\amin{\mathit{Am}}
\def\pmax{\mathit{PM}}
\def\pmin{\mathit{Pm}}

\begin{proof}
		Let $\loops = \{ \alpha_0, \alpha_1, \ldots, \alpha_{h-1} \}$, and 
		denote by $v_i$ the first node of $\alpha_i$ in its representation as
		a cyclic sequence of nodes.
		%
		%
		For all $i= 0,\ldots,h-1$, let $\pi_i$ a (possibly empty) path that starts
		in the last node of $\alpha_{i}$ and ends in the first node of
		$\alpha_{(i+1)\!\!\!\!\!\mod n}$.
		Since $\loops$ is connected, it is possible to find such paths.
		Let $A_i$ 
		be the color difference matrix of $\alpha_i$, and let
		$P_i$ 
		be the color difference matrix of $\pi_i$.
		Moreover, let $(c_0,c_1,\ldots,c_{n-1})$ be the non-negative integers such
		that $\frac{\sum_{l=0}^{h-1} c_l A_{l}}{\sum_{h=0}^{h-1} c_l n_l} = L$.
		Then, we define the matrix $Z=\sum_{l=0}^{h-1} c_l A_{l}$.
		Finally, let $n_i$ the number of edges in $\alpha_i$ and $m_i$
		the number of edges in $\pi_i$.
		At this point, we define $n = \sum_{i=0}^{h-1} c_i \cdot n_i$ and $m =
		\sum_{i=0}^{h-1} m_i$.

		In order to construct a path with color limit $L$, we reason as follows.
		Since in general the loops in $\loops$ do not share a node
		with each other,
		to move from $\alpha_i$ to $\alpha_{i+1}$,
		we have to pay a price, represented by 
		the color difference matrix of $\pi_i$.
		In order to make this price disappear in the long-run,
		we traverse the loops $\alpha_i$ an increasing number of times:
		in the first round, we traverse it $c_i$ times,
		in the second round, $2 c_i$ times, and so on.
		Formally, the construction is iterative and at every round $i > 0$
		we add, to the already constructed path, the cycle $\rho_i$ defined by
		$$\rho_i = \alpha_0^{i c_0} \, \pi_0 \alpha_1^{i c_1} \, \pi_1
		\ldots \alpha_{h-1}^{i c_{h-1}} \, \pi_{h-1}.$$
		Note that the cycle $\rho_i$ starts and ends at node $v_0$
		and contains $m+ i \cdot n$ edges.
		       %
		The required infinite path is then
		$\rho = \rho_1 \rho_2 \ldots \rho_i \ldots$.
		We now show that this path has color limit $L$.
		For all $i>0$, let $l_i = \sum_{j=1}^{i} |\rho_j| =
		\sum_{j=1}^{i} (m+i \cdot n)$, so that $\rho^{\leq l_i} = \rho_1 \ldots \rho_i$.
		We can easily observe that for every $i > 1$, it holds
		$l_{i+1}-l_{i} = m+i\cdot n = m+(i-1)\cdot n +n = l_{i}-l_{i-1}+n$.

		Let $\amax_{j,j',i}$ (resp., $\amin_{j,j',i}$) be the maximum (resp., minimum) of
		the $(j,j')$-color difference among the prefixes of $\alpha_i$, i.e. $\amax_{j,j',i} =
		max\{ \diff_{j,j'}(\alpha_i^{\leq t}) \mid 1 \leq t \leq n_i \}$ (resp.,
		i.e. $\amin_{j,j',i} = min\{ \diff_{j,j'}(\alpha_i^{\leq t}) \mid 1 \leq t \leq
		n_i \}$).
		Moreover, let $\amax_{j,j'} = \sum_{i=0}^{h-1} c_i \cdot \amax_{j,j',i}$ (resp.,
		$\amin_{j,j'} = \sum_{i=0}^{h-1} c_i \cdot \amin_{j,j',i}$).
		Similarly, let $\pmax_{j,j',i}$ (resp., $\pmin_{j,j',i}$) be the maximum (resp.,
		minimum) of the $(j,j')$-color difference among the prefixes of $\pi_i$, precisely
		$\pmax_{j,j',i} = max\{ \diff_{j,j'}(\pi_i^{\leq t}) \mid 1 \leq t \leq m_i
		\}$ (resp., $\pmax_{j,j',i} \allowbreak = \allowbreak min\{
		\diff_{j,j',d}(\pi_i^{\leq t}) \allowbreak \mid \allowbreak 1 \leq t \leq
		m_i \}$).
		Moreover, let $\pmax_{j,j'} = \sum_{i=0}^{h-1} \pmax_{j,j',i}$ (resp., $\pmin_{j,j'}
		= \sum_{i=0}^{h-1} \pmin_{j,j',i}$).

		At this point, we are able to derive the following two inequalities
		regarding the $(j,j')$-color difference along $\rho_i$.
		\begin{align*}
		(1) & \;\; \pmin_{j,j'} + i \cdot \amin_{j,j'} &=& \sum_{i=0}^{h-1} \pmin_{j,j',i}
		      + \sum_{i=0}^{h-1} i \cdot c_i \cdot \amin_{j,j',i} 
                    &\leq& \diff_{j,j'}(\rho_i^{\leq k}) \\
	           & & \leq &
		   \sum_{i=0}^{h-1} \pmax_{j,j',i} + \sum_{i=0}^{h-1} i\cdot c_i \cdot \amax_{j,j',i} 
                    &=& \pmax_{j,j'}+i\cdot \amax_{j,j'}.\\
		(2) & \;\; \pmin_{j,j'} + i \cdot Z_{j,j'} & \leq & \diff_{j,j'}(\rho_i) 
                    &\leq& \pmax_{j,j'} + i \cdot Z_{j,j'}.
		\end{align*}
		Thus, in the infinite path $\rho$ at each step $t \in [l_r, l_{r+1})$, we
		have that the $(j,j')$-color difference has module
		\begin{align*}
		  \diff_{j,j'}(\rho^{\leq t}) &\leq \sum_{i=1}^r i \cdot Z_{j,j'} +
		  i \cdot \pmax_{j,j'} + |\diff_{j,j'}(\rho_{i+1}^{\leq t-l_r})| \\
		  &\leq	\sum_{i=1}^r i \cdot Z_{j,j'} + i \cdot \pmax_{j,j'}
		 + \pmax_{j,j'} + (i+1) \amax_{j,j'} \\
		&= \sum_{i=1}^r i \cdot Z_{j,j'} + (i+1) (\pmax_{j,j'} + \amax_{j,j'}).
		\end{align*}

		\begin{align*}
		  \diff_{j,j'}(\rho^{\leq t}) &\geq \sum_{i=1}^r i \cdot Z_{j,j'} +
		  i \cdot \pmin_{j,j'} + |\diff_{j,j'}(\rho_{i+1}^{\leq t-l_r})| \\
		  &\geq	\sum_{i=1}^r i \cdot Z_{j,j'} + i \cdot \pmin_{j,j'}
		+ \pmin_{j,j'} + (i+1) \amin_{j,j'} \\
		&= \sum_{i=1}^r i \cdot Z_{j,j'} + (i+1) (\pmin_{j,j'} + \amin_{j,j'}).
		\end{align*}

		Note that, for all $i>1$, it holds that
			$(\pmin_{j,j'} + \amin_{j,j'}) + i \cdot Z_{j,j'} +
			\min \{ \diff_{j,j'}(\rho^{\leq t})
			\mid t\in [l_{r-1}, l_r) \} \leq \min
			\{ \diff_{j,j'}(\rho^{\leq t}) \mid t\in [l_{r}, l_{r+1}) \}
			\leq \max \{ \diff_{j,j'}(\rho^{\leq t})
			\mid t\in [l_{r}, l_{r+1}) \} \leq
			(\pmax_{j,j'} + \amax_{j,j'}) + i\cdot Z_{j,j'} \allowbreak + \allowbreak
		  	\max \{\diff_{j,j'}(\rho^{\leq t}) \mid t\in [l_{r-1}, l_{r}) \}$.
		So, applying Lemma~\ref{lmm:1G} to $a_n = b_n$, $k=n$,
		$c = \pmin_{j,j'} + \amin_{j,j'}$,
		$c'=Z_{j,j'}$, $c''=\pmax_{j,j'} + \amax_{j,j'}$ and $x_i = l_i$,
		we obtain that
		$\lim_{k \to +\infty} \frac{\diff_{j,j'}(\rho^{\leq t})}{k}
		=\frac{Z_{j,j'}}{n}= L_{j,j'}$.
		%
\end{proof}

\noindent
The following theorem characterizes the existence of an infinite
path with a given color limit in terms of a property of the simple
loops in the graph.
It is an immediate consequence of Lemma~\ref{lmm:colorLimitSet} and Lemma~\ref{lmm:1H}.

\begin{theorem}{} \label{thm:1H}
  Let $G$ be a $k$-colored graph, there exists an infinite path
  with color limit $L$ iff there exists a connected set of
  simple loops having a n.l.c. with ratio $L$.
\end{theorem}

\noindent
Finally, the following lemma links the color-limit-$L$ system with
the existence of a set of simple loops with the desired property.

\begin{lemma}{} \label{lmm:colorLimitInPStepA}
  There exists a set of simple loops in $G$ with an n.l.c. of ratio $L$ iff
  the color-limit-$L$ system for $G$ is feasible.
\end{lemma}

\begin{proof}
%
	[only if] Assume that $\loops$ is a set of simple loops having
	an n.l.c. with ratio $L$. Let $c_{\sigma}$ be the coefficient associated
	with the loop $\sigma \in \loops$. We construct a vector
	$x\in \reals^m$ that satisfies the color-limit-$L$ system.
	First, define $h(e,\sigma)$ as $1$ if
	the edge $e$ is in $\sigma$, and $0$ otherwise.
      Then, we set $x_e = \sum_{\sigma \in \loops} c_{\sigma} h(e,\sigma)$.
	Considering that, for all $\sigma\in \loops$ and $v\in V$,
	it holds that $\sum_{e \in _vE} h(e,\sigma) = \sum_{e\in E_v} h(e,\sigma)$,
        it is a matter of algebra to show that $x$ satisfies the color-limit-$L$
         system.

	[if] If the system is feasible, since it has integer coefficients,
        it has to have a rational solution.
        Moreover, all constraints are either equalities or inequalities
        of the type $a^T x \sim 0$, for $\sim\in\{ >, \geq\}$.
        Therefore, if $x$ is a solution then $c x$ is also a solution,
        for all $c > 0$. Accordingly, if the system has a rational solution,
        it also has an integer solution $x\in \ints^m$.
        Due to the constraints (3), such solution must be non-negative.
        So, in fact $x \in \nats^m$.

	Then, we consider each component $x_e$ of $x$
        as the number of times the edge $e$
	is used in a set of loops, and we use $x$ to construct such set with
	an iterative algorithm.
	At the first step, we set $x^1 = x$,
        we take a non-zero component $x^1_e$ of $x^1$,
	we start constructing a loop with the edge $e$, and then we subtract
	a unit from $x^1_e$ to remember that we used it. Next, we look for
	another non-zero component $x^1_{e'}$ such that $e'$ exits from the node
	$e$ enters in. It is possible to show that the edge $e'$ can always be found.
	Then, we add $e'$ to the loop and we subtract a unit from $x^1_{e'}$.
	We continue looking for edges $e'$ with $x^1_{e'}>0$ and exiting from
	the last node added to the loop, until we close a loop, i.e.,
	until the last edge added enters in the node the first edge $e$ exits from.
	After constructing a loop, we have a residual vector $x^2$ for the next step.
	If such vector is not zero, we construct another loop, and so on until
	the residual vector is zero.
In the end we have a set of (not necessarily simple) loops. 
Using inductive properties propagated through the steps of the algorithm, 
it is possible to show that the set of loops has an n.l.c. with ratio $L$.
        Finally, we decompose those loops
	in simple loops with the algorithm of Lemma 1 of \cite{BFMM09}, and we obtain
	the thesis.
%
\end{proof}

\noindent
Now, Lemma~\ref{lmm:colorLimitInPStep1} is an immediate corollary of
Theorem~\ref{thm:1H} and Lemma~\ref{lmm:colorLimitInPStepA}.

\section{Conclusions}
\label{sec:conclusion}

We have studied two-player games on colored graphs where 
the objective of player $0$ is the construction of a
balanced, bounded, or frequency-$f$ path. 
We have proved that deciding whether there exists a winning strategy 
for this player is a \conp-complete problem. 
Moreover, we have studied the one-player version of the games 
with the frequency-$f$ goal and shown that it is solvable in polynomial time.


An open natural question arising in this framework is the following:
if on a colored graph, or game, there is no bounded nor balanced path,
what is the ``most balanced path'' one can achieve?
This problem requires the definition of an appropriate order relation on 
color sequences, defining when a path is ``more balanced'' than another.

%

\noindent
\textbf{Acknowledgment.}
We thank Marcin Jurdzi\'nski for useful comments on the fre\-quen\-cy-$f$ problem.

\bibliographystyle{eptcs}
\bibliography{biblio}


\end{document}